\documentclass[aps,prl,twocolumn,showpacs,amsmath,amssymb]{revtex4-1}% APS journal style

\usepackage{graphicx}    % Include figure files
\usepackage{latexsym}    % 
\usepackage{color}       % 
\usepackage{dcolumn}     % Align table columns on decimal point
\usepackage{bm}          % bold math
\usepackage{amssymb}     % for math
\usepackage{hyperref}    % add hypertext capabilities
\expandafter\let\csname equation*\endcsname\relax
\expandafter\let\csname endequation*\endcsname\relax
\usepackage{amsmath}
\usepackage{ulem}

\usepackage{graphics}
\usepackage{epsfig}
\usepackage{verbatim}

\begin{document}

\title{Dynamo enhancement and mode selection triggered by high magnetic permeability}

\author{S Kreuzahler$^1$, Y Ponty$^2$, N Plihon$^3$, H Homann$^2$ and R Grauer$^1$}
\address{$^1$ Institut f\"ur Theoretische Physik I, Ruhr-Universit\"at Bochum, 44780 Bochum, Germany}
\address{$^2$ Universit\'e de la C\^ote d'Azur, CNRS, Observatoire de la C\^ote d'Azur, 
B. P. 4229 06304 Nice Cedex 4 , France}
\address{$^3$ Univ Lyon, ENS de Lyon, Univ Claude Bernard Lyon 1, CNRS,
Laboratoire de Physique, F-69342 Lyon, France}

\begin{abstract}
 We present results from consistent dynamo simulations, where the
 electrically conducting and incompressible flow inside a cylinder
 vessel is forced by moving impellers numerically implemented
 by a penalization method. The numerical scheme models jumps of
 magnetic permeability for the solid impellers, resembling
 various configurations tested experimentally in the von-Karman Sodium
 experiment. The most striking experimental observations are reproduced in
 our set of simulations. In particular, we report on the existence of a time averaged
 axisymmetric dynamo mode, self-consistently generated when the
 magnetic permeability of the impellers exceeds a threshold.  We
 describe a possible scenario involving both the turbulent flow in
 the vicinity of the impellers and the high magnetic permeability of
 the impellers.
\end{abstract}


% 47.27.ek= DNS;   :04450
% 47.11.Kb, 47.27.er = spectral methods; 
% 47.27.E- = turbulence simulation and modeling
% 47.65.-d 	Magnetohydrodynamics and electrohydrodynamics
% 47.11.-j = computational methods;  ? not found in base IAPS  (old pacs number ?) 

%\begin{keyword}
%Volume penalization \sep Pseudospectral method \sep Moving boundaries \sep von K\'arm\'an flow}
%\end{keyword}

\maketitle

\textit{Introduction: }\label{sec:intro}
Nearly a century ago, Larmor suggested that the dynamo effect, an instability converting kinetic energy into magnetic energy, could be at the origin of most astrophysical magnetic fields. The experimental observation of the dynamo instability has been a long quest requiring careful flow optimization, and has only been achieved in the Riga~\cite{gailitis2001}, Karlsruhe~\cite{stieglitz2001} and von K\'arm\'an sodium (VKS)~\cite{paper:monchaux:2007} experiments. While the behavior of the two former experiments could be explained from computations using simplified flows, this is not the case for the VKS experiment - in which a strongly turbulent liquid sodium flow is driven by the counter-rotation of impellers fitted with blades in a cylindrical vessel. Two major puzzles in the understanding of the dynamo mechanism are still unanswered: (i) the dynamo instability was only observed in the presence of impellers having high magnetic permeability~\cite{paper:miralles:2013} and (ii) the time-averaged dynamo magnetic field in the saturated regime  has an axial dipolar structure~\cite{paper:boisson:2012}, while an equatorial dynamo dipole is expected from computations in the growing phase of the instability  using the time-averaged axisymmetric flow~\cite{paper:ravelet:2005}.

Several numerical models have been proposed to explain these features of the VKS dynamo in the framework of the kinematic dynamo problem. The decrease of dynamo onset when implementing ferromagnetic boundary conditions (FBC) was first observed on the equatorial dipole using the time-averaged flow~\cite{paper:gissinger:2008.1,paper:giesecke:2012}. Below dynamo onset and with ferromagnetic impellers, toroidal modes were shown to be the least damped modes by paramagnetic pumping~\cite{paper:giesecke:2012}. The excitation of an axial dipole dynamo mode was then investigated following a turbulent $\alpha-\omega$ mechanism~\cite{paper:petrelis:2007} involving an $\omega$-effect from the time-averaged flow shear layer and a mean-field $\alpha$-effect from coherent vortices created between the blades, according to a mechanism proposed by Parker~\cite{paper:Parker:1955}. The $\alpha$ tensor from these vortices was  later confirmed by hydrodynamic computations~\cite{paper:ravelet:2012}. However, vortices velocities  of the order of the impeller tip velocity were required to obtain an axial dipole dynamo in this simplified $\alpha-\omega$ framework~\cite{paper:laguerre:2008,paper:laguerre:2008erratum,paper:gissinger:2009}. These models implementing ad-hoc mean-field terms were extended with accurate FBC for the impellers~\cite{paper:giesecke:2010,paper:nore:2015}. The influence of FBC on the focusing of vortices between the blades and on the dynamo process ($\alpha-\omega$ or $\alpha^2$) was also investigated~\cite{paper:varela:2015,paper:varela:2017}. Recently a first numerical model implementing accurate FBC for the impellers and a realistic flow in the laminar regime showed a transition from an equatorial dipole to an axial dipole and a decrease of dynamo onset as the magnetic permeability increases~\cite{nore_direct_2016}.

In this letter, by solving a complete incompressible magnetohydrodynamics (MHD) system implementing turbulent flow driven by moving impellers (using a penalization method) and finite magnetic permeability jumps, we present detailed numerical dataset which reproduce observations from the VKS experiment. In light of these results, we propose a new complex scenario for the VKS dynamo  based on the interaction of the flow and the high permeability impellers.

\vspace{.2 cm}

%%%%%%%%%%%%%%%%%%%%%%%%%%%%%%
\textit{Numerical method: }\label{sec:num}
We consider the MHD equations
\begin{eqnarray}
\partial_t{\mathbf{u}} + (\mathbf{u} \cdot \nabla) \mathbf{u} &=& - \nabla p + \nu \nabla^2 \mathbf{u} + \nabla \times ( \mu_r^{-1} \mathbf{b} ) \times \mathbf{b},\label{NSeq}\\
\partial_t{\mathbf{b}} + (\mathbf{u} \cdot \nabla) \mathbf{b} &=& (\mathbf{b} \cdot \nabla) \mathbf{u} - \nabla \times \left[ \frac{1}{\mu_0 \sigma} \nabla \times ( \mu_r^{-1} \mathbf{b} ) \right],\label{Indeq}\\
\nabla \cdot \mathbf{u}  = 0&,&
\nabla \cdot \mathbf{b}  = 0\label{Incompeq}
\end{eqnarray}
with the velocity field $\mathbf{u}(\mathbf{x},t)$, the magnetic induction field 
$\mathbf{b}(\mathbf{x},t)$ (rescaled by $1/\sqrt{\rho \mu_0}$ as an Alfv\'enic velocity - $\rho$ being the density of the fluid), the pressure $p(\mathbf{x},t)$ and the kinematic viscosity $\nu$. 
The electrical conductivity $\sigma(\mathbf{x},t)$  and the relative magnetic permeability $\mu_r(\mathbf{x},t)$ may be inhomogeneous. In order to reproduce the experimental configurations~\cite{paper:miralles:2013}, 
the conductivity is kept constant ($\sigma(\mathbf{x},t) =\sigma$) while the magnetic permeability may have distinct values in the fluid ($\mu_r=1$) and in the solid impellers driving the flow ($\mu_r=1$ up to $16$).
The MHD equations are solved with a standard Fourier pseudo-spectral method on a regular Cartesian grid.
The fluid domain is restricted to a cylinder with embedded boundaries. The rotating solid objects and outer boundaries are modeled via a penalization method~\cite{paper:fadlun:2000}, which recently showed its ability to reproduce extended experimental results on von-K\'arm\'an water flows~\cite{paper:kreuzahler:2014}.  
To handle the magnetic permeability jump and to prevent Gibbs oscillations, a sharpened raised cosine filter~\cite{script:canuto:1991} 
was implemented for the computation of derivatives.
For simplicity, periodic boundary conditions are used for the magnetic induction.
The periodic simulation box has dimensions of $(3\pi)^3$ with $384^3$ grid points. 
The fluid domain is restricted to a cylinder of radius $R_c~=~3.0$ and height $6.0$.  
This leaves a significant volume of the simulation domain where the fluid is at rest, 
avoiding spurious periodic solutions of the magnetic field (the magnetic energy at the boundaries of the simulation domain is  $4.\ 10^{-4}$ lower than its maximum value).
The curved-blade impeller setup is similar to the VKS experiment~\cite{paper:ravelet:2005, paper:monchaux:2007}, consisting of a disk with radius $R_d = 0.75 R_c$ fitted with 8 blades of height $0.2 R_c$ (25 grid points) and curvature radius $0.9 R_c$. The thickness of the blades and the disk is $0.15$ 
(6 grid points). 
The distance between the inner faces of the impellers is $1.8 R_c$.  
In all simulations, both impellers rotate with opposite angular velocity $\Omega=1.5$ 
leading to kinetic and magnetic Reynolds numbers $Re=\Omega R_d R_c/\nu=10.125/\nu$ and $Rm=\mu_0 \sigma \Omega R_d R_c$  (the fluid is pushed by the convex side of the blades).
Features for laminar and turbulent flows are presented for four distinct values of $Re$, namely $Re \sim 1500, 1000, 500, 333$ (obtained by changing the viscosity~$\nu$).

In this letter, we focus on the influence of the relative magnetic permeability $\mu_r$ of the solid impellers on the dynamo instability. 
In the following, three symmetric configurations are discussed:
the whole impeller having the same permeability $\mu_r\geq 1$ (referred to as "full impeller"), 
disk with $\mu_r=1$ and blades with $\mu_r=16$ ("blades only") and the reciprocal case ("disk only"). Our set of simulations incorporate all important ingredients of the VKS experiment, namely (i) turbulent flows, (ii) material properties jumps, (iii) back reaction of the Lorentz force (allowing to reach saturation of the dynamo instability), which are self-consistently solved. 

\vspace{.2 cm}

%%%%%%%%%%%%%%%%%%%%%%%%%%%%%%
\textit{Simulation results: }\label{sec:numresult}

\begin{figure}[t!]
\includegraphics[width=1\columnwidth]{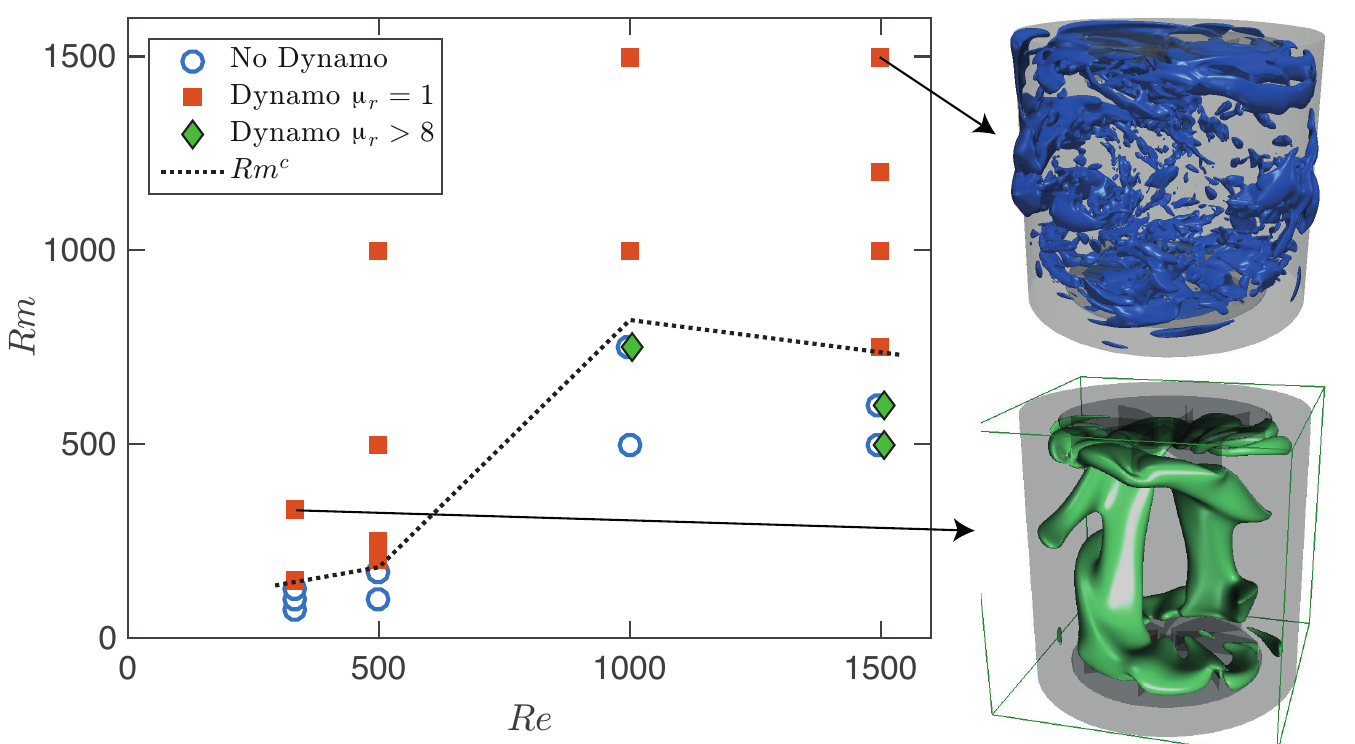}
\caption{($Re,Rm$) parameter space showing non-dynamo (blue empty circle) and dynamo runs (red square), and the dynamo onset (black dotted line) for $\mu_r=1$. Below dynamo onset at $\mu_r=1$, dynamos were observed for $\mu_r > 8$ (green diamonds). Right: Iso-contours of the time-averaged dynamo magnetic energy field in the growing phase for $Re = 333$ and  $Re = 1500$ ($Pm=1$).}
\label{fig:parameterspace}
\end{figure}

All numerical simulations are reported in the ($Re,Rm$) 
parameter space displayed in Fig.~\ref{fig:parameterspace} for the homogeneous situation ($\mu_r=1$ everywhere).
The black dotted line sketches the boundary between non-dynamo and dynamo runs, and shows that the critical magnetic Reynolds number $Rm^c$, above which a dynamo is observed, first increases as  turbulence of the von-Karman flows increases. The critical $Rm^c$ then saturates for turbulent flows ($Re>1000$). This is the confirmation, in fully self-consistent simulations including realistic boundary conditions, of an important result previously observed in simulations with periodic boundary conditions and different volume forcings~\cite{paper:ponty:2005,mininni_low_2005,paper:schekochihin:2007,ponty2011}. 
At low $Re$, the magnetic dynamo mode is an equatorial dipole, 
as expected for the mean-flow~\cite{paper:ravelet:2005,paper:gissinger:2008.1,paper:giesecke:2010}. 
At larger $Re$ values, the dynamo magnetic field is still mostly an equatorial dipole, but other azimuthal modes become significant (see iso-contours in Fig.~\ref{fig:parameterspace}).

Let us now focus on the influence of the relative permeability of the impellers $\mu_r$ for $Re=1500$, 
\textit{i.e.} in the turbulent regime. For $\mu_r = 1$, the dynamo onset is found around $Rm^c \sim 740$. Therefore, we will investigate further the $Rm=500$ case (see Fig.~\ref{fig:parameterspace}).

\begin{figure}
\includegraphics[width=.5\columnwidth]{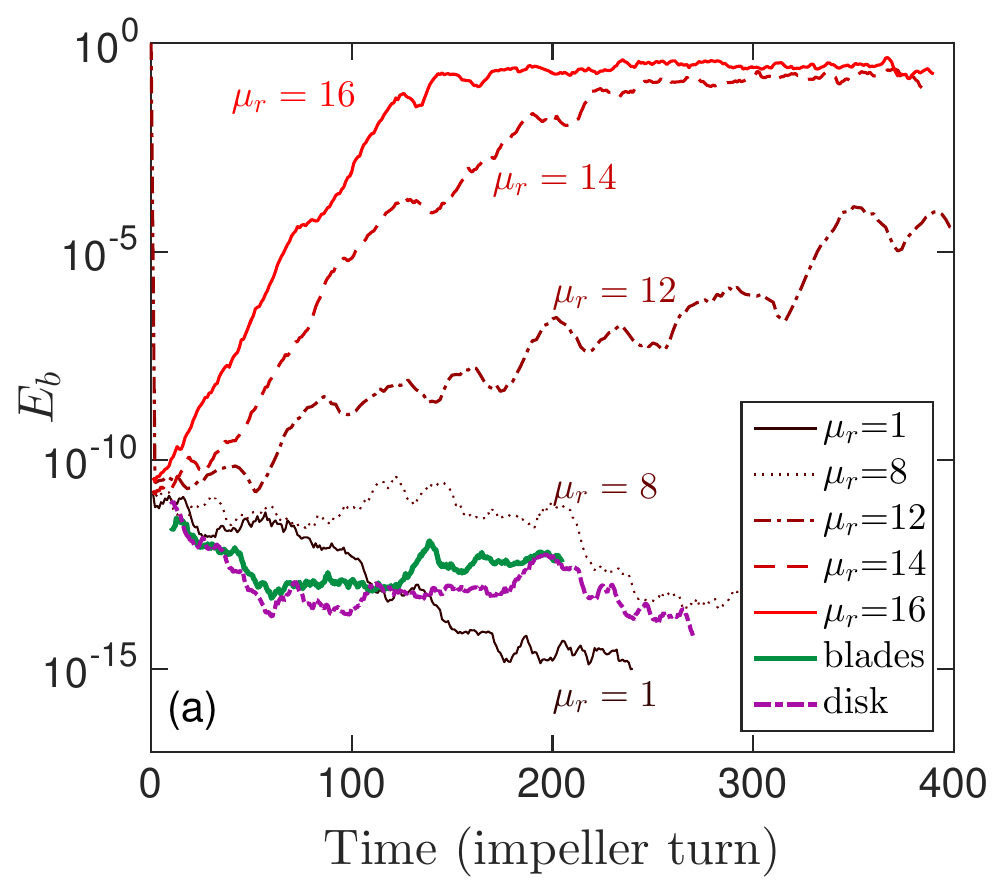}\includegraphics[width=.5\columnwidth]{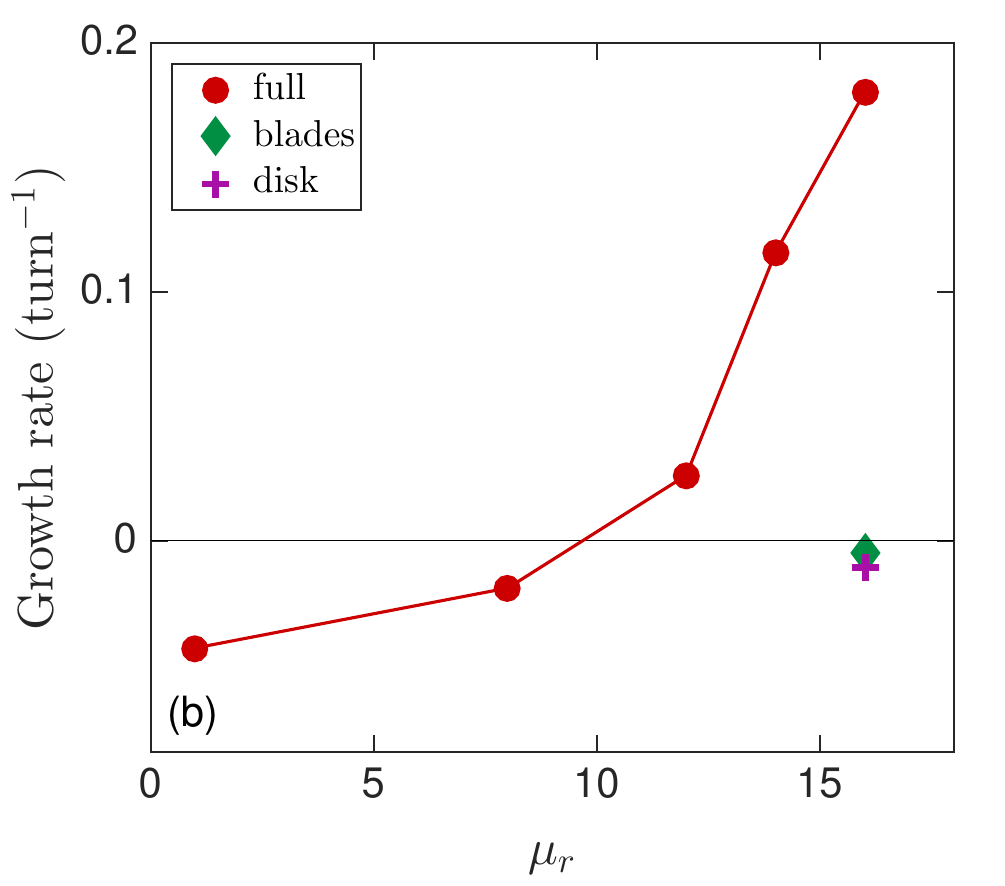}\\
\includegraphics[width=.5\columnwidth]{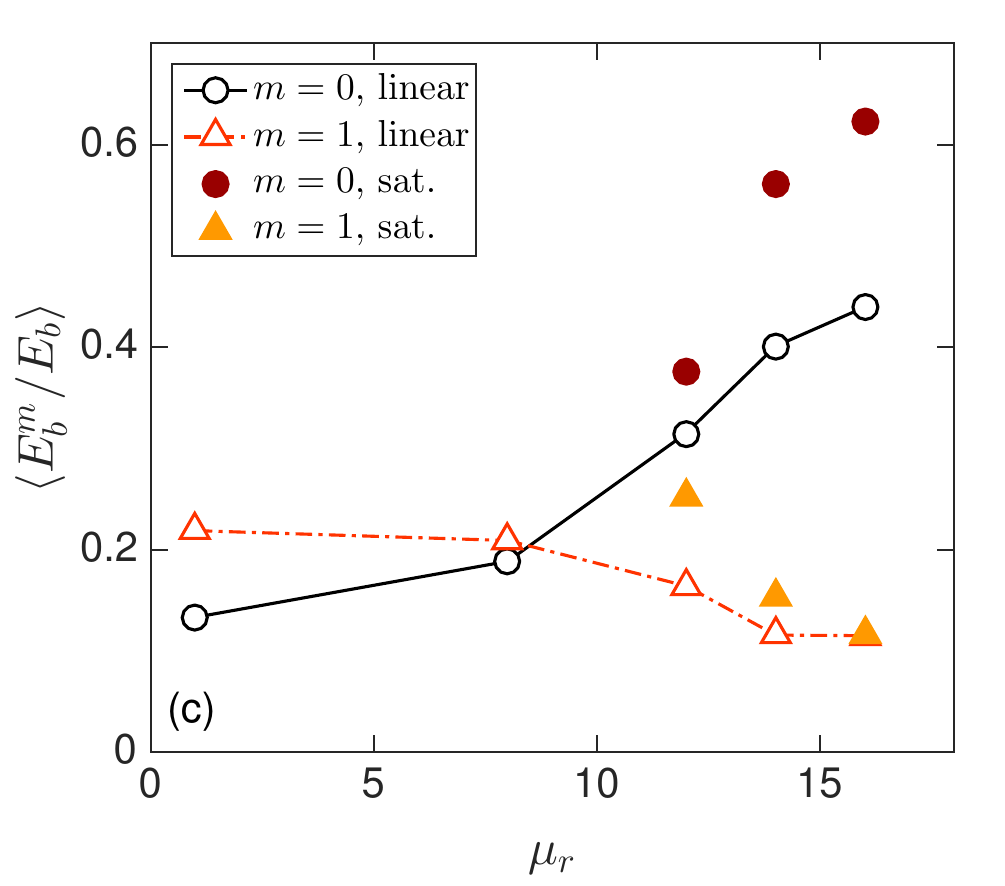}\includegraphics[width=.5\columnwidth]{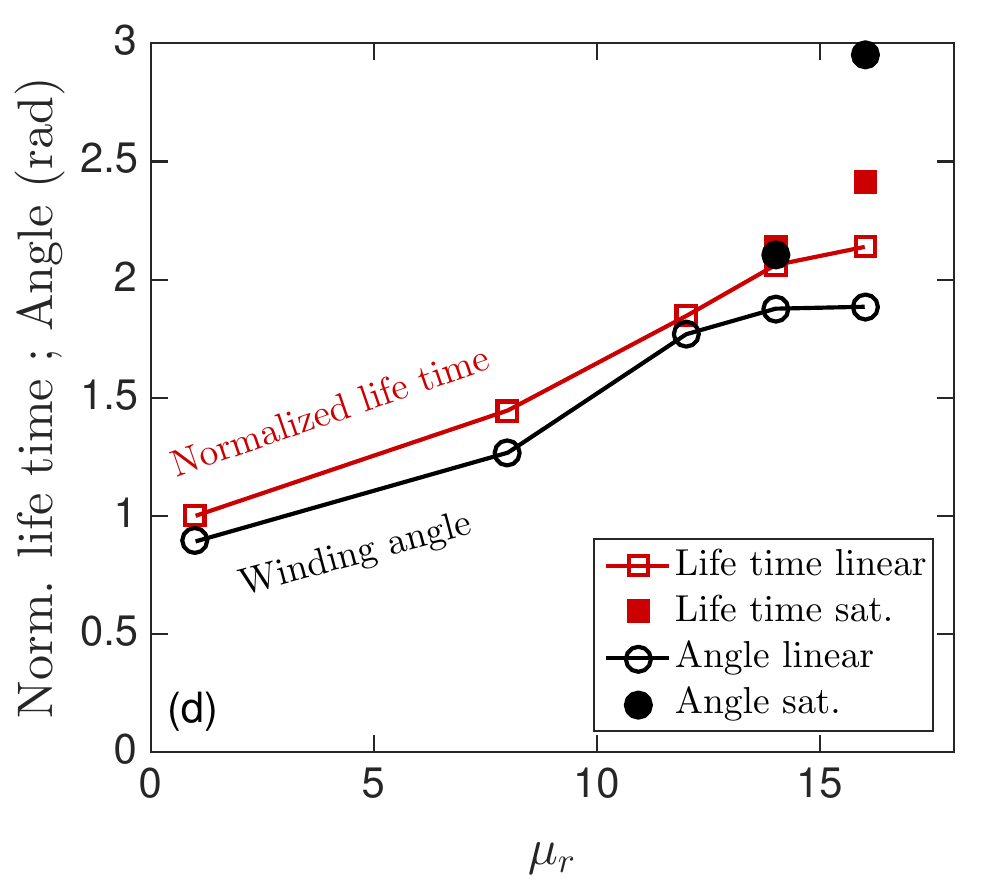}
\caption{(a) Time evolution of magnetic energy $E_b$, (b) dynamo growth rates, (c) dynamo energy ratio in modes $m=0$ and $m=1$, (d) normalized life time of magnetic field lines winding number in the vicinity of the impellers and as a function of $\mu_r$. See text for details.}
\label{fig:growthofmagneticfield}
\end{figure}

The magnetic energy is defined as \mbox{$E_b=\displaystyle\frac{1}{2V} \int_V \displaystyle\frac{|\mathbf{b}|^2}{\mu_r} dv$} where $V$ is the total volume. Fig.~\ref{fig:growthofmagneticfield}(a) displays its time evolution (in impeller turn units) for the three impeller configurations, and clearly shows that dynamo regimes are obtained by solely increasing  $\mu_r$~\cite{paper:gissinger:2008.1,paper:giesecke:2010,nore_direct_2016}. A similar behavior was also observed at $Re=1000, Rm=750$ and $Re=1500, Rm=600$. 
Fig.~\ref{fig:growthofmagneticfield}(b) shows the evolution of the dynamo growth rates 
(computed from the linear fit of the logarithmic time evolution of $E_b$ in the growing phase) 
as a function of $\mu_r$. For the "full impeller" configuration, dynamos are observed for $\mu_r \geq 12$.  Dynamo onset is lowered as $\mu_r$ increases as was previously extrapolated from magnetic induction measurements~\cite{paper:verhille:2010,paper:miralles:2013}.  
Another result similar to the VKS experiment is that the "disk only" and "blades only" configurations are no dynamos for $\mu_r = 16$ in our simulations~\cite{paper:miralles:2013} 
(though the set of dimensionless parameters differ from that of the experiment).

Based on a cylindrical coordinate system ($r,\phi,z$) within the fluid cylinder, further understanding is gained by investigating the azimuthal mode decomposition of the magnetic energy \mbox{$E_b^m = \displaystyle\frac{1}{2V} \iint \left[\mathbf{b}^m(r,z)\right]^2dr dz$}
with \mbox{$\mathbf{b}^m(r,z) = \displaystyle\int_0^{2\pi}\displaystyle\frac{\mathbf{b}(r,\phi,z)}{\sqrt{\mu_r(r,\phi,z)}}e^{i m\phi} d\phi$}. 
A similar mode decomposition is also computed for the kinetic energy.
 
\begin{figure}
\includegraphics[width=.48\columnwidth]{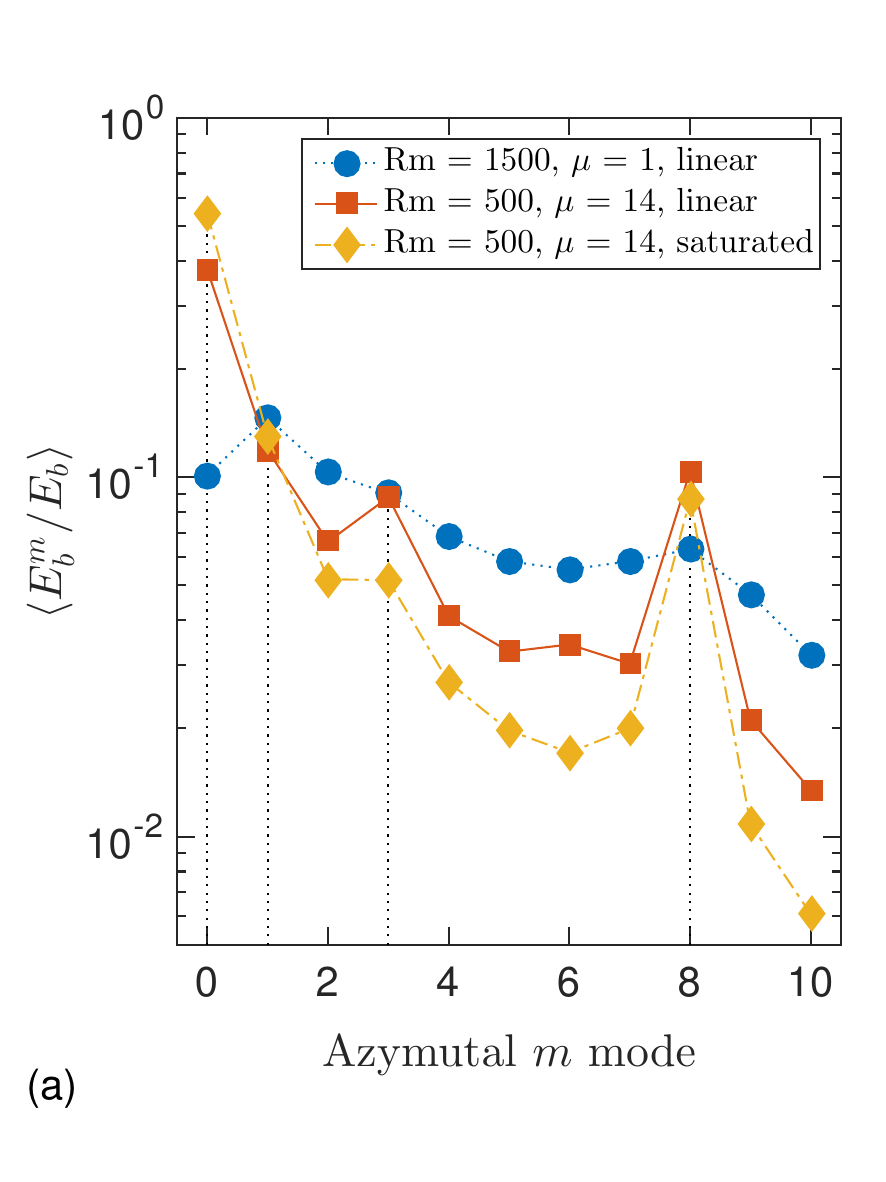}
\includegraphics[width=.5\columnwidth]{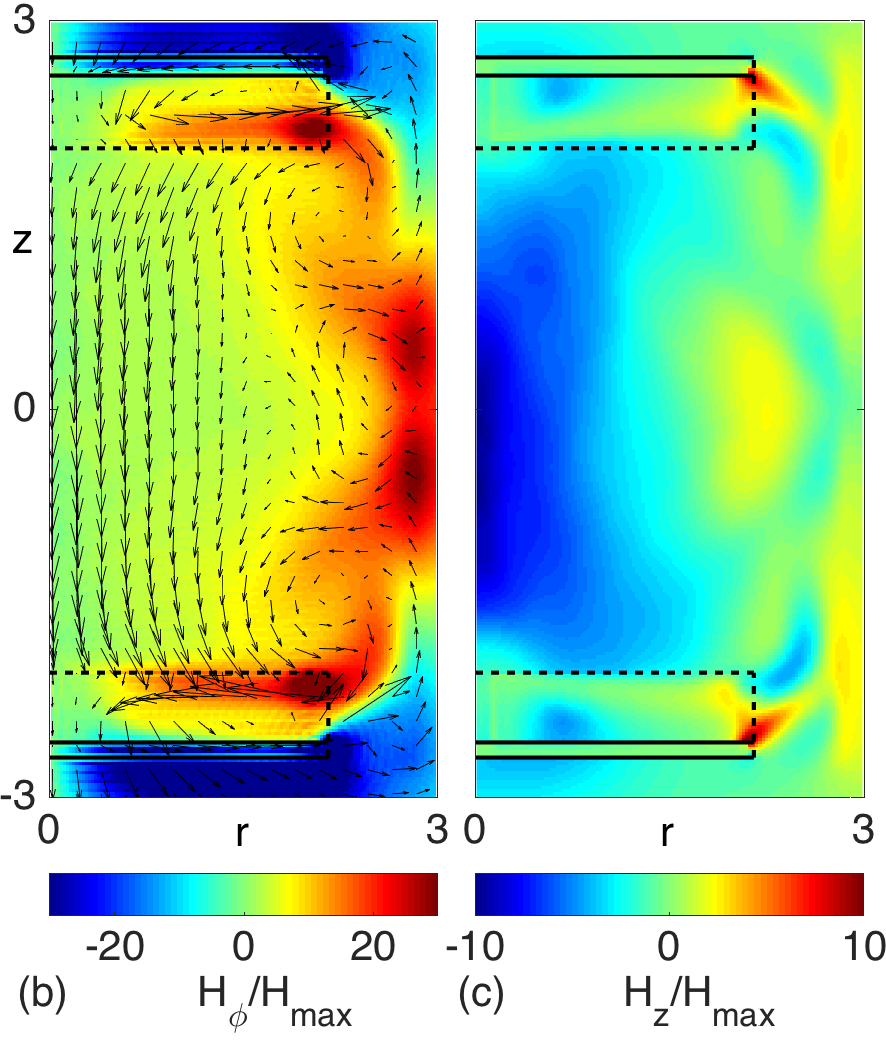}
\caption{(a) Azimuthal mode decomposition of magnetic energy, (b) azimuthally averaged $H_z$ and (c) $H_\phi$ components normalized to the maximum value in the linear regime for $Re=1500$, $Rm=500$, $\mu_r = 14$.  (b) Vectors show the polo\"\i dal dynamo field.}
\label{fig:energymode}
\end{figure}

The energy ratio $E_b^m/E_b$ for the leading $m=0$ and $m=1$ modes are displayed as a function of $\mu_r$ in Fig.~\ref{fig:growthofmagneticfield}(c), 
once again for $Re=1500$, $Rm=500$. An equatorial dipole (dominant $m=1$ mode) is observed for the decaying runs ($\mu_r<10$), while an axial dipole ($m=0$ mode) is clearly observed for the dynamo runs ($\mu_r \geq 12$). Large values of $\mu_r$ thus favor an axial dipole. It is important to note that the energy in the axisymmetric $m=0$ mode is stronger in the saturated regime than in the growing (or linear) phase.  This is further evidenced in Fig.~\ref{fig:energymode}(a) which reports the energy spectra of the dynamos observed at $Rm=1500$, $\mu_r=1$ (referred to as turbulent dynamo (TD) in the remaining) and $Rm=500$, $\mu_r = 14$ (referred to as high magnetic permeability-enhanced dynamo (PED)), both at $Re=1500$. The TD is dominated by an equatorial dipole, with large amounts of energy also present in the $m=0,2$ and 3 modes. 
The PED, on the other hand, is clearly an axial dipole.    
For both TD and PED, the $m=3$ mode of the magnetic energy is associated with large vortices created in the shear-layer and already observed in fully-turbulent water experiments~\cite{paper:delatorre:2007,cortet_normalized_2009} -- thus showing the ability of our simulations to reproduce turbulent features. In our simulations, these vortices are  damped by the Lorentz force in the saturated regime.
The $m=8$ mode is associated with eight vortices excited between the impeller blades~\cite{paper:kreuzahler:2014}, which are also slightly damped in the saturated regime. 
As a partial conclusion, our set of simulations thus reproduce three important features relative to the influence of $\mu_r$ in the VKS experiment: (i) large values of $\mu_r$ decrease the dynamo onset in the "full impeller" configuration (ii) hybrid configurations are no dynamo, (iii) the presence of ferromagnetic impellers leads to a transition from an equatorial to an axial dipole.

The azimuthally averaged axisymmetric components of the PED $\mathbf{H} = \mathbf{b}/\mu_r$ field are displayed in Fig.~\ref{fig:energymode}(b) and (c). Most of the energy of the azimuthal component is located close to the high permeability impellers, as observed in the VKS experiment~\cite{paper:boisson:2012} and recent numerical simulations~\cite{nore_direct_2016}. The axial component $H_z$ is strong and homogeneous in the flow bulk, with opposite polarity close to the outer radius, as in the VKS experiment~\cite{paper:boisson:2012}. A close investigation of the magnetic energy evolutions reported in Fig.~\ref{fig:growthofmagneticfield}(a) shows oscillations both in the linear and the saturated phases; these oscillations correspond to polarity changes, as in the VKS experiment~\cite{paper:ravelet:2008.1}.  These findings will be reported in details elsewhere.

\vspace{.2 cm}

\begin{figure}
\includegraphics[width=.9\columnwidth]{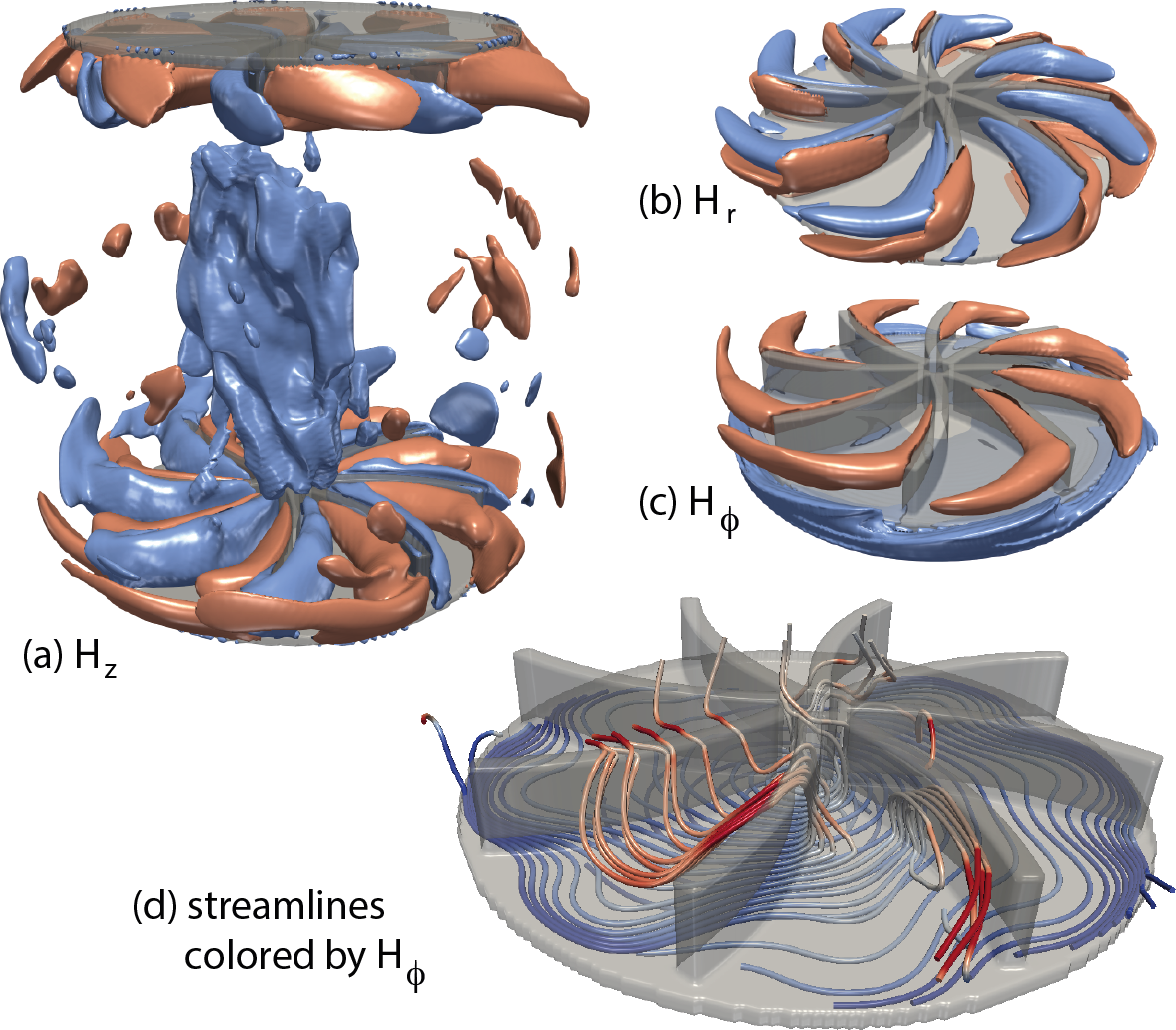}
\caption{(a) $H_z$, (b) $H_r$, (c) $H_\phi$ iso-amplitudes contours (red: positive value, blue: opposite negative value) and (d) magnetic streamlines colored by $H_\phi$ value. $Re=1500$, $Rm=500$ and $\mu_r=14$.}
\label{fig:3Dvisu}
\end{figure}

%%%%%%%%%%%%%%%%%%%%%%%%%%%%%%
\textit{Magnetic streamline topology and high magnetic permeability scenario: }\label{sec:topology}
The $m=1$ dominated time-averaged structure of the TD  is well understood from kinematic computations using the time-averaged von-Karman flow. On the other hand, the observation of an axial dipole for the PED is linked to the ferromagnetic nature of the impellers. 
To gain a better understanding, 3D visualizations of the iso-contours of the $\mathbf{H}$ components of the dynamo field are provided in Fig.~\ref{fig:3Dvisu}, averaged over every impeller turn in the growing phase. Fig.~\ref{fig:3Dvisu}(a) shows a strong and uniform $H_z$ in the bulk, which is the signature of the $m=0$ dynamo. In the vicinity of the impellers, the structure of $H_z$ is much more complex  with alternate polarities of $H_z$ on either sides of the blades. A similar pattern is observed for $H_r$ in Fig.~\ref{fig:3Dvisu}(b). Regarding the $H_\phi$ component, a first observation is  that  the strong value of the $H_\phi$ component inside the disk has the same sign as behind the impeller. This can be understood from the flow behind the impellers, where a strong shear layer produces a strong $\omega$-effect (this shear layer is not present in Ref.~\cite{nore_direct_2016}), and from continuity conditions of the electromagnetic fields leading to magnetic field line refraction at high  $\mu_r$~\cite{paper:giesecke:2010,paper:giesecke:2012,nore_direct_2016}. 
A second observation is that, slightly above the impeller, the strong shear layer above the blades (of opposite sign as compared to behind the disk) leads to an $H_\phi$ component with the opposite sign of that behind the impeller. While the $H_\phi$ component changes sign throughout the impeller, the $H_z$ component remains of the same sign (see Fig.\ref{fig:energymode}(c)); the complex geometry thus rules out the symmetry arguments provided in~\cite{herault_optimum_2014} for a simpler geometry. 
The features reported for each of the $\mathbf{H}$ components are summarized in the streamlines visualization reported in Fig.~\ref{fig:3Dvisu}(d) and lead to a third observation:  towards the core of the flow, the magnetic connection between the impeller and the flow occurs through the blades.
The effect of the coherent vortices (and thus of the $\alpha$-effect introduced in~\cite{paper:petrelis:2007}) is clearly observed on the streamlines shown in Fig.~\ref{fig:3Dvisu}(d).

We now propose to explain the polo\"\i dal to toro\"\i dal conversion occurring from the complex coupling between the flow features in the vicinity of the impellers and their ferromagnetic nature, by introducing new quantitative diagnostics. At each 3D snapshot recorded along the simulation, one thousand seed points are randomly set within a test cylinder close to the impellers.  The magnetic ($\mathbf{H}$) streamlines starting from theses points are integrated until they exit the test cylinder. 
Two scalars are then extracted. The first one, referred to as "life time", is the averaged integration time of the streamline remaining inside the test cylinder: the larger it is, the more efficient is the trapping of magnetic field lines in the vicinity of the impellers. The second one, referred to as "winding angle", is the averaged azimuthal angle experienced by the streamline within the test cylinder: this can be viewed as a effective polo\"\i dal to toro\"\i dal magnetic field conversion.
These quantities, plotted in Fig.~\ref{fig:growthofmagneticfield}(d), strongly increase with $\mu_r$. The interplay between the flow features in the vicinity of the impellers and the high magnetic permeability thus leads to an effective $\omega$-effect, significantly enhanced by $\mu_r$.
The evolution of these global quantities strongly supports a localized scenario for the dynamo process: (i) toro\"\i dal to polo\"\i dal conversion from  an $\alpha$-effect linked to coherent vortices between the blades~\cite{paper:petrelis:2007,paper:laguerre:2008,paper:varela:2017} and (ii) polo\"\i dal to toro\"\i dal conversion, strongly enhanced by the high magnetic permeability of the impellers, occurring in the vicinity of the impellers. In this scenario, both conversion processes are located close to the coherent vortices within the blades, contrary to the $\alpha-\omega$ mechanism proposed in Ref.~\cite{paper:petrelis:2007}, which had detrimental effects~\cite{paper:laguerre:2008}. Moreover, the scenario demonstrated here strongly depends upon the magnetic permeability of the impellers.

\vspace{.2 cm}

%%%%%%%%%%%%%%%%%%%%%%%%%%%%%%
%\textit{High permeability scenario:}\label{sec:scenario}
%\input{sections/scenario.tex}
%%%%%%%%%%%%%%%%%%%%%%%%%%%%%%

\textit{Conclusion:}
A set of numerical simulations is presented where the MHD equations, including flow drive from impellers rotation (via a penalization term) and jumps of magnetic permeability $\mu_r$, are self-consistently solved in a geometry close to that of the VKS experiment. This set simultaneously reproduces the most important experimental observations in a self-consistent treatment without additional ad-hoc terms: (i) large $\mu_r$ values decrease the dynamo onset, (ii) large $\mu_r$ values lead to a transition from an equatorial to an axial dipole, (iii) the blades and the disk of the impellers are equally important in the dynamo generation. We demonstrate a  scenario where an effective $\mu_r$-enhanced $\omega$-effect occurs in the vicinity of the impellers and takes into account the complex coupling of the transport of the magnetic field by the flow and the ferromagnetic structure. This polo\"\i dal to toro\"\i dal conversion process adds up to a polo\"\i dal to toro\"\i dal conversion from an $\alpha$-effect linked to coherent vortices between the blades, leading finally to an axial dipole dynamo mode.
%\vspace{.1 cm}

\textit{Acknowledgments:}
Parts of this research were supported by Research Unit FOR 1048,
project B2, and the French Agence Nationale de la Recherche under
grant ANR-11-BLAN-045, projet SiCoMHD.
Access to the IBM BlueGene/P computer JUGENE at the FZ J\"ulich was
made available through the project HBO40.
Computer time was also provided by GENCI in the IDRIS/CINES/TGCC national french facilities,
the Mesocentre SIGAMM, hosted by the Observatoire de la C\^ote d'Azur and 
the CICADA computer facilities hosted by University of Nice-Sophia.

\bibliographystyle{apsrev4-1}

\bibliography{referencesNPv3}

\end{document}